\documentstyle[prd,aps,epsfig,twocolumn,floats]{revtex}
\begin{document}
\draft
\title{The transformations of the electromagnetic potentials
under translations}
\author{Bernd A. Berg (berg@hep.fsu.edu) }
\address{Department of Physics, The Florida State University,
  Tallahassee, FL 32306}
\date{ June 4, 2000 -- hep-th/0006029; revised June 12, 2000}
\maketitle
\begin{abstract}
I consider infinitesimal translations
$x'^{\alpha}=x^{\alpha}+\delta x^{\alpha}$
and demand that Noether's approach gives a symmetric electromagnetic
energy-momentum tensor as it is required for gravitational sources.
This argument determines the transformations of the electromagnetic
potentials under infinitesimal translations to be $A'_{\gamma} (x') =
A_{\gamma}(x)+\partial_{\gamma} [\delta x_{\beta}\, A^{\beta}(x)]$,
which differs from the usually assumed invariance 
$A'_{\gamma} (x') = A_{\gamma}(x)$, by the gauge transformation
$\partial_{\gamma} [\delta x_{\beta}\, A^{\beta}(x)]$.

\end{abstract}

\narrowtext


In relativistic field theory it is well known, and often referred to
as Noether's theorem~\cite{Noether}, that each independent 
infinitesimal symmetry implies a conserved current with an associated 
constant or motion.  Here we are interested in the symmetry under 
infinitesimal translations
\begin{equation} \label{translations}
x'^{\, \alpha} = x^{\alpha} + \delta x^{\alpha}
\end{equation}
for which Noether's theorem yields the conserved currents of the
energy-momentum tensor with the associated constants of motion 
being energy and momenta. However, the energy-momentum tensor 
$T^{\alpha\beta}$
obtained in this way from the electromagnetic Lagrangian
\begin{equation} \label{LED}
 {\cal L} = - {1\over 16\pi}\, F_{\alpha\beta}\, F^{\alpha\beta}
~~~{\rm with}~~ F^{\alpha\beta} = \partial^{\alpha} A^{\beta} 
  - \partial^{\beta} A^{\alpha} 
\end{equation}
is not symmetric, whereas everyone believes that the correct
result ought to be symmetric. The standard procedure~\cite{Jackson}
is to add a total divergence such that the final result becomes
the desired symmetric tensor $\theta^{\alpha\beta}$. While this
procedure is acceptable within electrodynamics, it becomes 
questionable as soon as one is concerned about gravity.
The electromagnetic energy-momentum distributions of
$T^{\alpha\beta}$ and $\theta^{\alpha\beta}$ differ and this changes
the implied gravitational force. This is in principle
observable~\cite{FeynmanII}, although in practice presumably not,
unless someone identifies suitable cosmological field distributions.
On the theory side, in gravity the symmetry transformations of general 
covariance yield the symmetric energy-momentum tensor as source of the 
gravitational field, see for instance~\cite{LaLi02,WeiG}, and this is 
presumably the strongest evidence underlining that the energy-momentum 
distribution of $\theta^{\alpha\beta}$ is the correct one. 

In view of the arguments in favor of the symmetric energy-momentum
tensor it is astonishing that Noether's theorem leads to a tensor
$T^{\alpha\beta}$ with an obviously incorrect energy-momentum
distribution. In particularly, one should bear in mind that the
derivation of $T^{\alpha\beta}$ relies only on the symmetry
transformations of a four vector under translations
\begin{equation} \label{4vector_trans}
 A'_{\gamma} (x') = A_{\gamma}(x)
\end{equation}
and on factoring out the {\it local} translation $\delta x^{\alpha}$.
Due to the locality of the procedure, it is hard to imagine that it
could lead to an incorrect energy momentum distribution when the
fundamental assumptions are sound.

In this letter I give a simple solution to the problem. The
non-symmetric energy momentum tensor $T^{\alpha\beta}$ is obtained
under the assumption that $A_{\gamma}$ transforms as a four
vector~(\ref{4vector_trans}) under translations~(\ref{translations}).
Due to the gauge invariance
of the electromagnetic Lagrangian~(\ref{LED}) this does not have to
be the case. We may allow for more general transformations which
differ from~(\ref{4vector_trans}) by gauge transformations, {\it i.e.}
\begin{equation} \label{gauge_trans_ansatz}
A'_{\gamma} (x')=A_{\gamma}(x)+\partial_{\gamma}\,\Lambda (x)\ .
\end{equation}
That nature may employ such a transformation behavior instead
of~(\ref{4vector_trans}) is not entirely a surprise.
On the quantum level the electromagnetic fields rely on 
superpositions of massless creation and annihilation operators and 
Weinberg~\cite{WeinbergI} points out to us that such fields do not 
allow for representations of the (proper) Lorentz group, but only 
for transformations which differ from those by a gauge transformation.
Therefore, it is quite natural to conjecture that nature uses
gauge transformation also for translations. Repeating the
arguments of Noether's theorem with the 
ansatz~(\ref{gauge_trans_ansatz}) and requesting a symmetric 
energy-momentum tensor leads to the unique solution
\begin{equation} \label{gauge_trans}
A'_{\gamma} (x')=A_{\gamma}(x)+\partial_{\gamma}\,
[ \delta x_{\beta}\, A^{\beta}(x) ]
\end{equation}
which is conjectured to be the transformation law realized by
nature for infinitesimal translations of electromagnetic
potentials. The remainder of the paper is devoted to the
derivation of this equation. Up to
some notational changes and adaptions to the case at hand, my
arguments follow closely chapter~1 of Bogoliubov and
Shirkov~\cite{BS}.

First, let us quickly recall how relativistic field equations are 
derived from the action principle. The action is a four dimensional 
integral over a scalar Lagrangian density
\begin{equation} \label{action}
{\cal A} = \int d^4x\, {\cal L} 
(\psi_k,\, \partial_{\alpha}\psi_k)
\end{equation}
and, therefore, by itself a scalar under the connected part of the
Lorentz group. Variations of the fields are defined as functions
\begin{equation} \label{fvariations}
\delta\psi_k (x) = \psi'_k (x) - \psi_k (x)\, 
\end{equation}
which are non-zero for some localized space-time region. The action
is required to vanish under such variations
$$ 0 = \delta {\cal A} = $$
\begin{equation} \label{Lvar1}
 \sum_k \int d^4x\, \left[
(\delta\psi_k)\, {\partial {\cal L}\over \partial \psi_k} + 
(\delta\, \partial_{\alpha}\psi_k)\, {\partial {\cal L}\over 
\partial (\partial_{\alpha}\psi_k)} \right]\ . 
\end{equation}
Integration by parts gives 
\begin{equation} \label{Avariation}
0 = \sum_k \int d^4x\, (\delta\psi_k)\, \left[ {\partial 
 {\cal L}\over \partial \psi_k} - \partial_{\alpha}\, {\partial 
 {\cal L}\over \partial (\partial_{\alpha}\psi_k)} \right]\,,
\end{equation}
where we used that the surface terms vanish.
As the variations $\delta \psi_k$ are independent, the integrand
in (\ref{Avariation}) has to vanish for each $k$ and we arrive at 
the Euler-Lagrange equations
\begin{equation} \label{EulerLagrange}
{\partial {\cal L}\over \partial \psi_k} - \partial_{\alpha}\,
{\partial {\cal L}\over \partial (\partial_{\alpha}\psi_k)} = 0
\end{equation}
for relativistic fields. For the electrodynamic Lagran\-gian~(\ref{LED})
they yield $ \partial_{\alpha}\, F^{\alpha\beta} = 0 $.

Noether's theorem applies to transformations of the 
coordinates for which the transformations of the field functions are
also known. Such transformations constitute a symmetry of the theory
when the corresponding variation of the action vanishes. The theorem
states that to each such symmetry a combination of the field functions 
exists which defines a conserved current. For this purpose we introduce,
in addition to (\ref{fvariations}), a second type of variations which
combines space-time and their corresponding field variations
\begin{equation} \label{cvariations}
\overline{\delta} \psi_k (x) = \psi'_k (x') - \psi_k (x)\, . 
\end{equation}
Using (note $\delta x^{\alpha}\,\partial_{\alpha}\psi_k' = 
\delta x^{\alpha}\,\partial_{\alpha}\psi_k$ because $\delta^2$
variations disappear)
$$ \psi'_k (x') = \psi'_k (x) + 
\delta x^{\alpha}\ \partial_{\alpha} \psi_k (x) $$
we find a relation between the variations (\ref{cvariations}) and 
(\ref{fvariations})
\begin{equation} \label{rvar}
\overline{\delta} \psi_k (x) = \delta\psi_k (x) +
\delta x^{\alpha}\ \partial_{\alpha} \psi_k (x)\, .
\end{equation}
For a scalar field $\psi$ (as well as for ordinary four vector
fields) symmetry under translations means
\begin{equation} \label{ctrans}
\overline{\delta} \psi (x) = \psi' (x') - \psi (x) = 0\, . 
\end{equation}
But for the electromagnetic potentials we
allow~(\ref{gauge_trans_ansatz})
\begin{equation} \label{vtrans}
\overline{\delta} A_{\gamma} (x) = A'_{\gamma} (x') - A_{\gamma} (x)
 = \partial_{\gamma} \Lambda (x)\,.
\end{equation}
With these symmetries equation (\ref{rvar}) reduces for a scalar 
field to
\begin{equation} \label{rctrans}
\delta\psi = - \delta x^{\alpha}\ \partial_{\alpha} \psi (x)
\end{equation}
and for the electromagnetic potentials to
\begin{equation} \label{rvtrans}
\delta A_{\gamma} = \partial_{\gamma} \Lambda (x) 
- \delta x^{\alpha}\ \partial_{\alpha} A_{\gamma} (x)\ .
\end{equation}
As the Lagrange density is a scalar, we get for its combined
variation~(\ref{cvariations})
\begin{equation} \label{cvarL}
0 = \overline{\delta} {\cal L} = {\cal L}' (x') - {\cal L} (x) = 
\delta {\cal L}+ \delta x^{\alpha}\ \partial_{\alpha} {\cal L}
\end{equation}
where besides~(\ref{ctrans}) we used the relation~(\ref{rvar}).
Our aim is to factor an over-all variation $\delta x^{\alpha}$ 
out. For $\delta {\cal L}$ we proceed as in equation~(\ref{Lvar1}),
where the fields $\psi_k$ are now replaced by the gauge
potentials $A_{\gamma}$ 
$$ \delta {\cal L} =  (\delta A_{\gamma})\, {\partial 
{\cal L} \over \partial A_{\gamma}} + (\delta \partial_{\alpha} 
A_{\gamma})\, {\partial {\cal L} \over \partial 
(\partial_{\alpha} A_{\gamma})}\,.$$
Using the Euler-Lagrange equation (\ref{EulerLagrange}) to
eliminate $\partial {\cal L} / \partial A_{\gamma}$, we get
(the calculation remains valid in our case where ${\cal L}$ does
not depend on $A_{\gamma}$) 
$$ \delta {\cal L} =  (\delta A_{\gamma})\, 
\partial_{\alpha} {\partial {\cal L} \over \partial 
(\partial_{\alpha} A_{\gamma})} + (\delta \partial_{\alpha} A_{\gamma})
\, {\partial {\cal L} \over \partial (\partial_{\alpha} A_{\gamma})} $$
$$ = \partial_{\alpha} \left[ (\delta A_{\gamma})\, {\partial {\cal L}
 \over \partial (\partial_{\alpha} A_{\gamma})} \right]\, .$$
Let us collect all terms which contribute  to
$\overline{\delta}{\cal L}$ in equation~(\ref{cvarL}).
For this, note that
$\partial_{\beta}\delta x^{\alpha}=0$ holds for all combinations of 
indices $\alpha$, $\beta$. (Namely, for $\alpha=\beta$ we are led to 
$\delta 1=0$ and for $\beta \ne \alpha$ the variations $\delta x^{\alpha}$ 
are then independent of the coordinates $x^{\beta}$). We find 
$$ 0=\overline{\delta} {\cal L} = \partial_{\alpha}\, 
\left[ (\delta A_{\gamma})\, {\partial {\cal L}\over \partial 
(\partial_{\alpha} A_{\gamma)}} + \delta x^{\alpha}\ {\cal L}
\right] = $$
$$ \partial_{\alpha}\, \left[ \left( \partial_{\gamma} \Lambda(x) -
(\delta x_{\beta})\, \partial^{\beta} A_{\gamma} \right) \,
{\partial {\cal L}\over \partial (\partial_{\alpha} A_{\gamma})}
+ g^{\alpha\beta}\, \delta x_{\beta}\, {\cal L} \right]  $$
where equation~(\ref{rvtrans}) was used for the last step. To be
able to factor $\delta x_{\beta}$ out of the bracket, one has to
request
\begin{equation} \label{translation_gauge}
 \Lambda (x) = \delta x_{\beta}\, B^{\beta}(x)
\end{equation}
where $B^{\beta}(x)$ is a not yet determined potential field. With this
we get
$$ 0 = \delta x_{\beta}\,
\partial_{\alpha}\, \left[ ( \partial^{\beta} A_{\gamma} -
 \partial_{\gamma} B^{\beta} )\, {\partial {\cal L}\over 
 \partial (\partial_{\alpha} A_{\gamma})}
- g^{\alpha\beta}\, {\cal L} \right]\,. $$
As the variations $\delta x_{\beta}$ are independent, the 
energy-momentum tensor 
\begin{equation} \label{e-mtensor}
\theta^{\alpha\beta} = {\partial {\cal L}\over \partial 
(\partial_{\alpha} A_{\gamma})}\ (\partial^{\beta} A_{\gamma} -
 \partial_{\gamma} B^{\beta}) - g^{\alpha\beta}\, {\cal L} 
\end{equation}
gives the conserved currents
\begin{equation} \label{e-mcurrent}
\partial_{\alpha}\, \theta^{\alpha\beta} = 0\, .
\end{equation}
Let us demand that the energy-momentum tensor~(\ref{e-mtensor}) is 
symmetric.  This leads to the requirement
\begin{equation}
B^{\beta}(x) = A^{\beta}(x)
\end{equation}
for which 
\begin{equation} \label{e-stensor}
\theta^{\alpha\beta} = {1\over 4\pi} \left(
F^{\alpha\gamma}\, F_{\gamma}^{\,\ \beta} + {1\over 4} g^{\alpha\beta}\,
F_{\gamma\delta}\,F^{\gamma\delta} \right) 
\end{equation}
is symmetric because of
$$ F^{\alpha\gamma}\, F_{\gamma}^{\,\ \beta} = 
   F^{\beta\gamma}\, F_{\gamma}^{\,\ \alpha}\ .  $$
Indeed, equation~(\ref{e-stensor}) is the symmetric tensor of
the textbooks~\cite{Jackson}. It differs from other versions
of~(\ref{e-mtensor}) by total divergencies.

In conclusion, I have presented an argument in favor of the
transformation behavior~(\ref{gauge_trans}) and it appears that
the question~\cite{FeynmanII} of the energy-momentum distribution 
of the electromagnetic field may finally be put at rest with
the expected result. Noether's theorem alone has no predictive 
power about whether the energy-momentum tensor is the symmetric
or not, but
evrything is consistent in the sense that a transformation law
exists which gives the tensor~(\ref{e-stensor}). In addition,
when symmetry of the tensor is assumed the result for 
$\theta^{\alpha\beta}$ is unique.
It is shown in a forthcoming paper that this approach 
works also for non-abelian gauge theories.
\hfil\break

\centerline{\bf Note added} \medskip

After posting this manuscript Prof. Jackiw kindly informed
me that my result is a special case of his work~\cite{Ja78},
see~\cite{JaMa80} for details. Prof. Hehl communicated that the
use of 1-Forms leads directly to a symmetric energy-momentum
tensor, see for instance~\cite{He95}.

\acknowledgments

This work was in part supported by the US Department of Energy under
contract DE-FG02-97ER41022.


\begin{thebibliography}{99}

\bibitem{Noether} E. Noether, Nachrichten Akademie der Wissenschaften,
G\"ottingen p.235 (1918).

\bibitem{Jackson} J.D. Jackson, {\it Classical Electrodynamics},
Second Edition, John Wiley \& Sons, 1975: Section 12.10.

\bibitem{FeynmanII} R.P. Feynman, R.B. Leighton and M. Sands, 
{\it The Feynman Lectures on Physics}, Volume~II, {\it Mainly 
Electromagnetism and Matter}, Addison-Wesley, 1964: chapter~27, 
section~4.

\bibitem{LaLi02} L.D. Landau and E.M. Lifshitz, {\it The Classical
Theory of Fields}, Course of Theoretical Physics~2, Butterworth
Heinemann, Oxford, 1995: Section 94.

\bibitem{WeiG} S. Weinberg, {\it Gravitation and Cosmology}, John
Wiley \& Sons, 1972: Chapter 12.

\bibitem{WeinbergI}  S. Weinberg, {\it The Quantum Theory of Fields I},
Cambridge University Press, 1998: Section 5.9.

\bibitem{BS} N.N. Bogoliubov and D.V. Shirkov, 
{\it Introduction to the Theory of Quantized Fields}, 
John Wiley \& Sons, 1959.

\bibitem{Ja78} R. Jackiw, Phys. Rev. Lett. 41, 1635 (1978).

\bibitem{JaMa80} R. Jackiw and N.S. Manton, Ann. Phys. 127, 257 (1980).

\bibitem{He95} F.W. Hehl, J.D. McCrea, E.W. Mielke, and Y. Ne'eman,
Phys. Rep. 258, 1 (1995).

\end{thebibliography}
\end{document}